# Laser written junctionless dual in-plane-gate thin-film transistors with AND Logic function

L.Q.Zhu, G.D.Wu, H.L.Zhang, J.M.Zhou, and Q.Wan[*]

*Ningbo Institute of Materials Technology and Engineering, Chinese Academy of Sciences, Ningbo 315201, People's Republic of China*

**Abstract**

A simple laser scribing process has been developed to fabricate low-voltage junctionless in-plane-gate thin-film transistors (TFTs) arrays without any mask and photolithography. Such junctionless TFTs feature that the channel and the source/drain electrodes are of the same indium-tin-oxide films without any intentional source/drain junction deposition process. Effective field-effect modulation of the drain current has been realized on such in-plane-gate device with a field-effect mobility of ~12.6 $cm^2$/Vs. At last, AND gate logic function was demonstrated on dual in-plane-gate device.

Keywords: Laser scribing, Junctionless TFTs, Dual in-plane-gate, AND logic.

[*] Corresponding author. Tel:/Fax:+86-574-86690355,
E-mail addresses: wanqing@nimte.ac.cn

# 1. Introduction

Field-effect transistors are deemed as the fundamental building blocks for the applications of the state-of-the art electronic devices. Moreover, dual gate oxide based thin film transistors (TFTs) have also been studied for a variety of electrical applications as pixel display driver or even logic circuit components.[1] In the conventional thin-film transistors (TFTs) fabrication processing, photolithography and shadow mask methods are inevitably adopted in order to form a patterned channel layer and source/drain electrodes. Recently, the concept of junctionless FET devices has been proposed and explored for simplifying the traditional junction-based device process. [2-4] Compared with conventional FETs, the unique feature of such transistors is that the channel doping is the same and comparable to that of the source and drain (S/D), therefore no sour/drain junction is needed, and the carrier transport is less sensitive to the channel interface. Such device structure can simplify the traditional junction-based fabrication process and save the cost by a great deal. However, the device process usually needs a complicate photolithography for patterning the heavily doped channel on SOI substrates, as is undesirable in terms of the applications in the low cost fabrication process. Therefore, the fabrication of junctionless thin-film transistors (TFTs) with simplified process is especially needed to be developed. By adapting a self-assembling process with only one nickel shadow mask, patterned channel layers have been achieved. Therefore, a coplanar junctionless TFTs have been obtained.[5, 6]

Laser patterning process offers the potential for inexpensive and flexible patterning.

More recently, a laser-delamination process has been applied to lift-off thin films on transparent substrate.[7] The irradiation of the interface between the film and the substrate with a short pulse of a laser gives rise to the localized heating, melting, and quenching near the interface, resulting in the separation of the film due to the thermal stress at the interface. Laser irradiation process has also been applied to fabricate the electronics devices. [8, 9] After a metal layer deposition on $SiO_2$/Si substrate, two laser irradiation processes have been adopted to fabricate the TFTs. [8] The first laser irradiation process spatially modulated through a shadow mask results in the sacrificial Al patterns. After the deposition of the ITO layer on, a second laser irradiation process is adopt to detach the Al/ITO stacks on the substrate, therefore the patterned ITO could be obtained. When depositing the semiconductor layer on, a bottom gated TFT is fabricated. Though the complicated photolithography and lift-off processes have been avoided, there are still two laser irradiation processes with the adoption of the shadow mask. The process could be simplified by depositing Al thin layers on ZTO/$SiO_2$/Si substrate. Only one pulsed laser beam irradiation step modulated through a shadow mask was needed. [9] The modulated laser beam selectively etched the Al film out of the underlying ZTO layer due to a thermoelastic force caused by rapid thermal expansion resulting form the pulsed laser irradiation. Take the isolated Al patterns as the source and drain electrode, while the bottom Si as gate electrode, a bottom gated TFT have been realized. Focused laser beam is also an important research tool to pattern the thin films. The creation of an effective localized heating zone results in a localized physical or chemical state transition. [10] The insulated graphene oxide (GO) nanosheets could be patterned in air by the focused laser beam results in the isolated GO island on quartz substrate due to the fact that the GO layer could be burned into CO or $CO_2$ when receiving laser irradiation. At the same time, the insulated graphene oxide (GO) nanosheets could also be reduced into conductive graphene layer when receiving laser irradiation in an inert environment. Therefore, the isolated conductive graphene pattern could also be obtained.

In this work, a focused laser beam process has been developed to write the TFTs arrays directly. Take the advantages of photolithography free and shadow mask free process, the developed novel laser scribing process is the simplest process to fabricate the junctionless TFTs. Junctionless TFTs arrays with an in-plane-gate structure gated by $SiO_2$-based solid electrolyte have been obtained. This maskless laser patterning technology could save much process time from mask alignment step. The most special feature of such TFTs is that the TFTs arrays are obtained through a laser scanning process free of any shadow mask and free of precise photolithography or alignment process. Moreover, the channel and source/drain electrodes are realized by a thin indium-tin-oxide (ITO) without any intentional source/drain junction. Effective field-effect modulation of the drain current has been realized on both the single in-plane-gate (SG) mode and the dual in-plane-gate (DG) mode. The dual gate (DG) ITO TFT has been characterized for advanced logic component.

## 2. Experimental details

The devices were fabricated on commercially purchased conducting ITO glass substrates, and the whole process was performed at room temperature, as schematically shown in Fig.1. First, a 100 nm Al thin films was deposited on ITO glass substrates by sputtering. Then, a 2-μm-thick $SiO_2$-based solid electrolyte film was deposited by plasma enhanced chemical vapor deposition (PECVD) using $SiH_4$ and $O_2$ as reactive gases, working as a EDL dielectrics.[11] Finally, a 30nm thick ITO films were deposited on the $SiO_2$-based solid electrolyte by sputtering ITO target in Ar ambient. The thickness of ITO films were deduced by using an ex situ spectroscopic ellipsometry (J.A.Woollam). The stack structure is shown in Fig.1 (a). Then the stacks were patterned by a laser scribing process, as shown in Fig.1 (b). The first laser scribing process would result in the isolation of the top ITO films, while the second scribing process would result in the isolation of both the top ITO films and the bottom ITO films. Fig.1 (c) illustrates the laser patterned ITO arrays with the dimension of 1mm×0.2mm. Fig.1(d) and Fig.1(e) shows a top-view optical image of the laser patterned junctionless single in-plane-gate (SG) and dual in-plane-gate (DG) TFTs in electrical measurement on a probe station, respectively. The nominal channel width (W) is 0.2mm, while the channel length (L) is 0.4mm and 0.2mm for SG TFT and DG TFT, respectively. Among them, DG mode is important for AND gate application, which needs simultaneous and independent sequential switching. The electrical characteristics of the junctionless DG or SG TFTs was measured with a keithley 4200 SCS semiconductor parameter analyzer at room temperature in the dark. The AND gating of our DG TFT in time domain can also be addressed by keithley 4200

SCS semiconductor parameter analyzer with source measurement units (SMUs). The capacitance-frequency measurement of the $SiO_2$-based solid electrolyte was performed using a Solartron 1260A Impedance/Gain-Phase Analyzer.

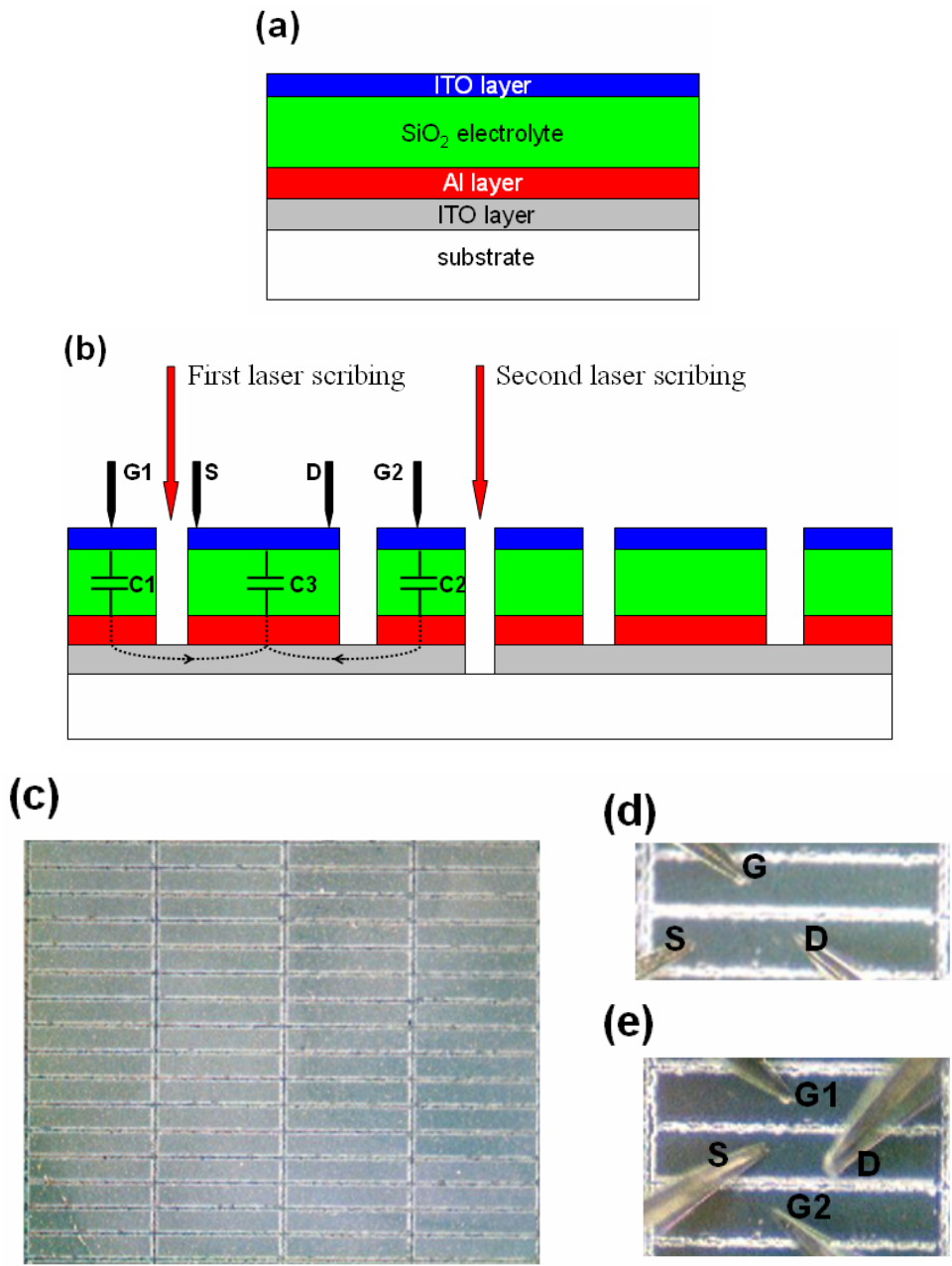

Fig.1 (a) The stack structure for preparing the junctionless thin film transistors (TFTs). (b) Schematic cross-sectional view of the laser patterned junctionless dual in-plane-gate TFTs structure. The first laser scribing process would result in the isolation of the top ITO films, while the second firing process would result in the isolation of both the top ITO films and the bottom ITO films. The three capacitors (C1, C2 and C3) are effectively coupled through bottom conducting ITO films. (c) The top view optical image of the junctionless TFTs arrays. (d) The top view optical image of the junctionless single in-plane-gate (SG) TFTs in electrical measurement on a probe station. (e) The top view optical image of the junctionless dual in-plane-gate (DG) TFTs in electrical measurement on a probe station.

## 3. Results and discussions

Capacitor of $SiO_2$-based solid electrolyte with two in-plane ITO electrodes is fabricated and measured. Fig.2 (a) shows the specific gate capacitance and the ionic conductivity of the $SiO_2$-based solid electrolyte in response to the frequency ranging from 1.0Hz to 1.0MHz, using an ITO/$SiO_2$/Al/ITO in-plane test structure (as shown in the upper parts in the inset of Fig.2 (a)). The values of the capacitance increase with the decreasing frequency. While the ion conductivities increase with the deceasing frequency. Fig.2 (b) also shows the gate leakage of such $SiO_2$ dielectric using an ITO/$SiO_2$/Al/ITO in-plane test structure. A low leakage current of below 0.2nA was measured, which guarantee that the transistor performance would not be affected by leakage. [6] Such behaviors strongly indicate that our PECVD-deposited $SiO_2$ film is an electronically insulating ion-conducting solid-electrolyte dielectric and is desirable for low-voltage operation of junctionless oxide based TFTs. [6, 12] The in-plane-gate capacitance at 1.0Hz shows a maximum value of 0.25μF/cm$^2$. As reported in Ref.[11], an equivalent circuit based on three capacitors (C1, C2, and C3) can be applied to understand the operation mechanism of DG device. The three capacitors are effectively coupled through bottom conducting films, as shown in the lower middle parts in the inset of Fig.2(a) and described in Fig.1(b). Thus, the channel current $I_{ds}$ controlled by G1, can be modulated by the secondary gate G2.

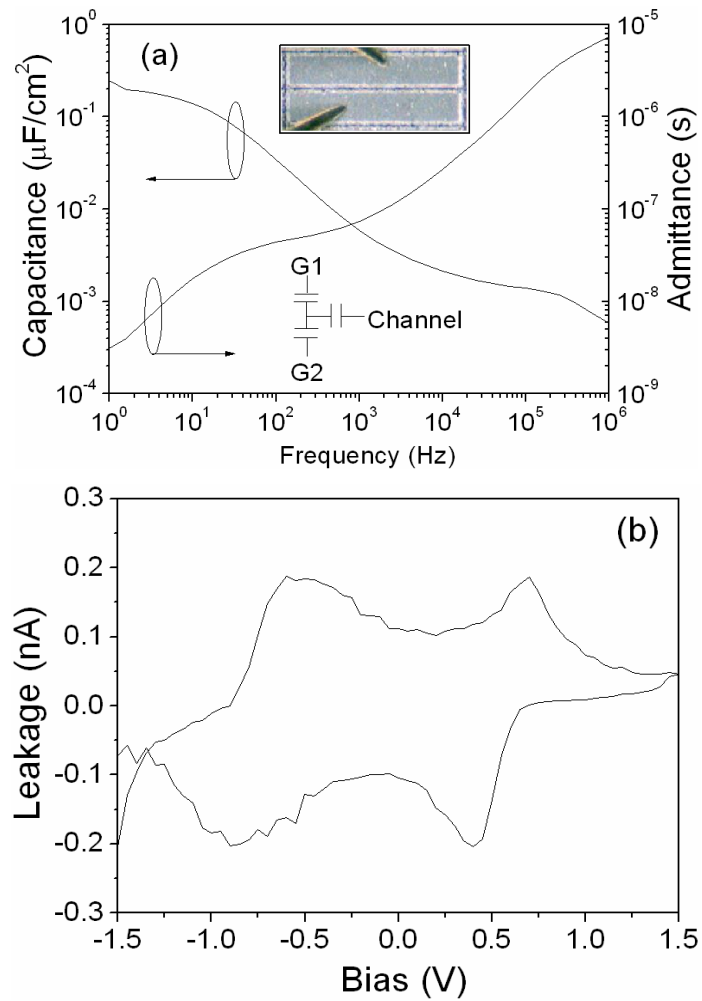

Fig.2 (a) Specific gate capacitance and the ionic conductivity of the SiO$_2$ electrolyte as a function of frequency. Inset: the top-view optical image of the ITO/SiO$_2$/Al/ITO in-plane test structure on a probe station. (b) The gate leakage of PECVD SiO$_2$ dielectric using an ITO/SiO$_2$/Al/ITO in-plane test structure.

The output characteristics of the laser patterned junctionless single in-plane-gate (SG) TFTs are shown in Fig.3(a). The nominal channel length is 0.4mm, while the channel width is 0.2mm. The V$_{gs}$ increases from -2V to 2V in 0.4V steps. At low V$_{ds}$, the drain current increases linearly with drain voltage, indicating that the device has a good ohmic contact. At the higher V$_{ds}$, the drain current gradually approaches a saturated value. These characteristics are quite similar to those of conventional field-effect transistors. With a low gate voltage of 2V, the saturation current is observed to be ~5μA at a drain voltage below 2V.

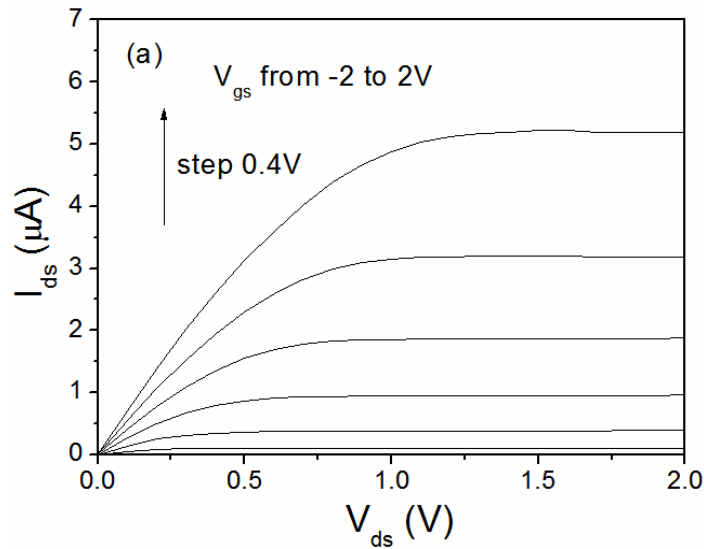

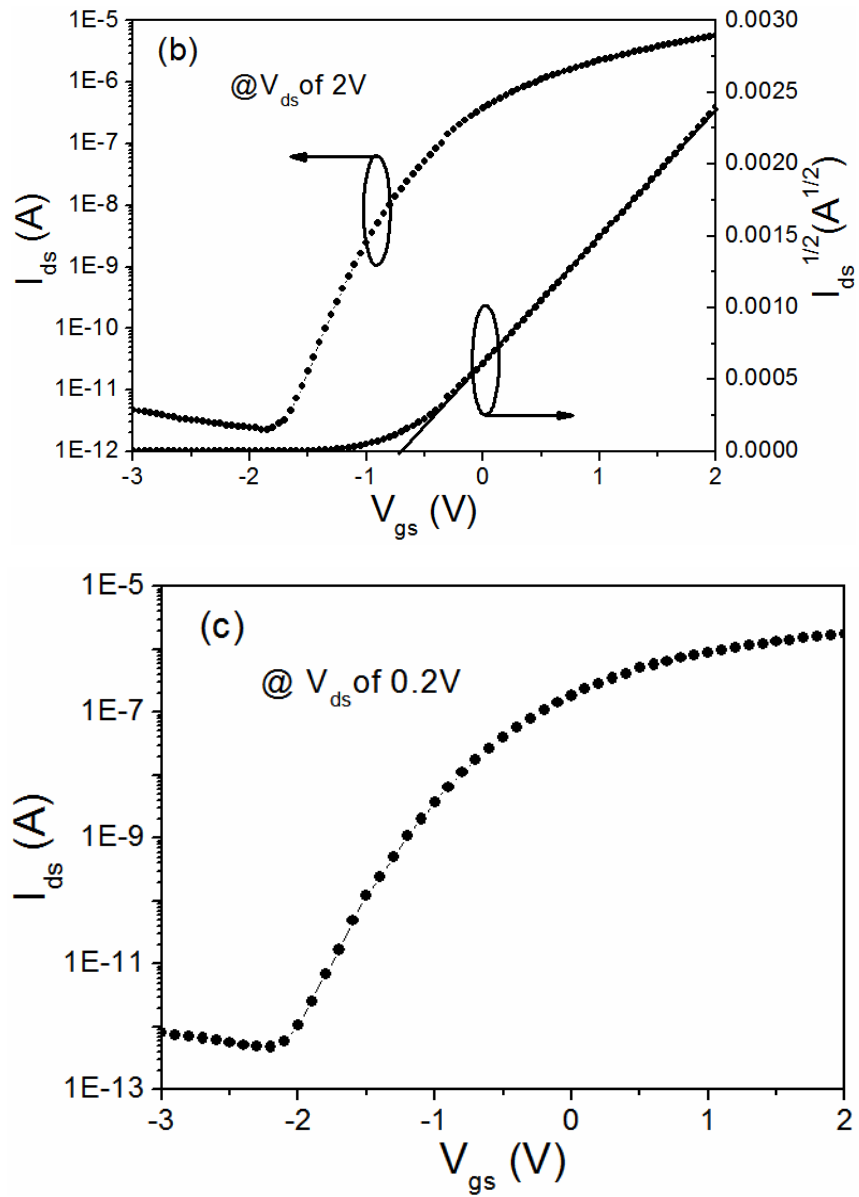

Fig.3 (a) Output characteristics of the laser patterned junctionless single in-plane-gate (SG) TFTs. (b) Transfer characteristics ($I_{ds}$ vs $V_{gs}$ and the square root of $I_{ds}$ vs $V_{gs}$) of the laser patterned junctionless single in-plane-gate (SG) TFTs at saturation mode, fixed at $V_{ds}$ of 2V respectively. (c) Transfer characteristics ($I_{ds}$ vs $V_{gs}$ and the square root of $I_{ds}$ vs $V_{gs}$) of the laser patterned junctionless single in-plane-gate (SG) TFTs at linear mode, fixed at $V_{ds}$ of 0.2V respectively.

Fig.3(b) and (c) shows the transfer characteristics of the laser patterned junctionless TFTs at linear and saturation mode, fixed at Vds of 0.2V and 2V, respectively. It can be found that $I_{ds}$ can be effectively modulated by $V_{gs}$. Moreover, the two curves show great accordance. For the transfer characteristics at saturation mode, the subthreshold swing (S) is found to be ~0.25V/dec. The drain current on/off ratio $I_{on}/I_{off}$ is determined to be $3\times10^6$. A threshold voltage $V_{th}$ of -0.7V was estimated by extrapolating the linear portion of the curves relating $I_{ds}^{1/2}$ and $V_{gs}$ to $I_{ds}^{1/2}=0$. The field-effect mobility (μ) in the saturation region can be extracted from the following equation:

$$I_{ds} = (\frac{WC_i\mu}{2L})(V_{gs} - V_{th})^2 \qquad (V_{ds} > V_{gs} - V_{th})$$

where L is the channel length, W is the channel width, and $C_i$ is the unit area capacitance of the dielectrics. A specific capacitance of 0.25μF/cm² at 1.0Hz was used to calculate the field-effect mobility. The channel width is ~0.2mm, while the channel length is ~0.4mm. The field-effect electron mobility is estimated to be ~12.6cm²/Vs.

To further test whether electrochemical doping effects is responsible for the observed electrical performance in the laser patterned TFTs, the pulse response measurement of $I_{ds}$ was performed at a constant $V_{ds}$ of 2V for SG TFTs. A square-shaped $V_{gs}$ with a pulsed amplitude of -3.0V and 2.0V were applied. Fig.4 illustrates the results. The junctionless TFT exhibits a reasonable reproducibility of the current response to the repeatedly pulsed $V_{gs}$. No on-current loss and a maintainable $I_{on}/I_{off}$ (~$10^6$) are obtained in response to the repeatedly pulsed $V_{gs}$ (~35 testing cycles). It was reported that $I_{ds}$ value would not return to its original value after gate scanning when there is the chemical doping or a chemical reaction at the interface [13]. Our results suggest that no chemical reaction is expected at the $SiO_2$-based solid electrolyte/ITO interface when the gate potential is biased.

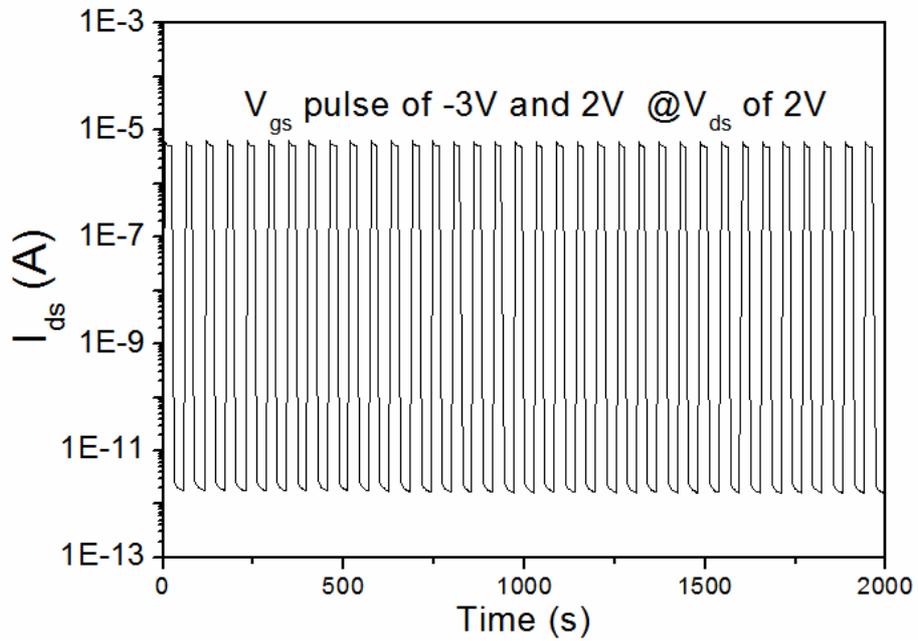

Fig.4 Transient response of the laser patterned junctionless in-plane-gate TFTs to square-shaped $V_{gs}$ pulses of -3V and 2V with a constant bias of $V_{ds}$=2V.

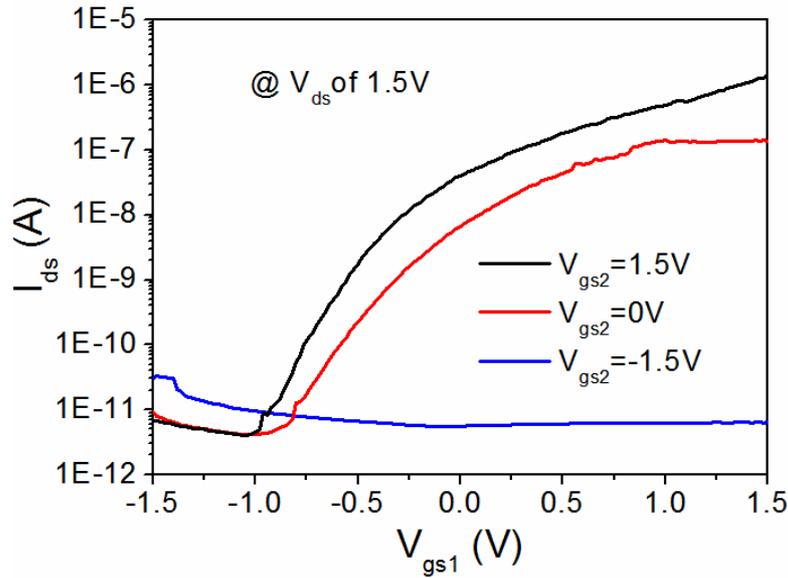

Fig.5 The transfer characteristics of the laser patterned junctionless dual in-plane-gate (DG) TFTs at saturation region, fixed at $V_{ds}$ of 1.5V.

Fig.5 shows the transfer characteristics of the laser patterned junctionless dual in-plane-gate (DG) TFTs at saturation region, fixed at $V_{ds}$ of 1.5V. The channel width is ~0.2mm, while the channel length is ~0.2mm. The drain currents are controlled by two gates (G1 and G2). The G1 bias sweeps from -1.5V to 1.5V with the fixed G2 bias set to be 1.5V, 0V and -1.5V, respectively. It can be obviously found from Fig.5 that the drain current can be effectively modulated by G1 bias with fixed G2 bias of 1.5V and 0V. The drain current can be turned on or turned off effectively with the drain current on/off ratio $I_{on}/I_{off}$ of above $10^5$ and above $10^4$ for G2 bias of 1.5V and 0 V, respectively. The subthreshold swing (S) is found to be ~0.25V/dec for G2 bias of 1.5V and ~0.33V/decade for G2 bias of 0V. A threshold voltage $V_{th}$ of -0.25V was estimated by extrapolating the linear portion of the curves relating $I_{ds}^{1/2}$ and $V_{gs}$ to $I_{ds}^{1/2}=0$ for G2 bias of 1.5V and 0V. Considering that the DG TFTs use two parallel capacitors serially connected to channel, the field-effect electron mobility is estimated to be ~3.8cm$^2$/Vs for G2 bias of 1.5V. All the results here indicate that the laser patterned in-plane-gate junctionless TFTs could operate under the low voltage of 1.5V. Interestingly, the drain current can not be turned on effectively under the G2 bias of -1.5V. These results would be meaningful for the logic application.

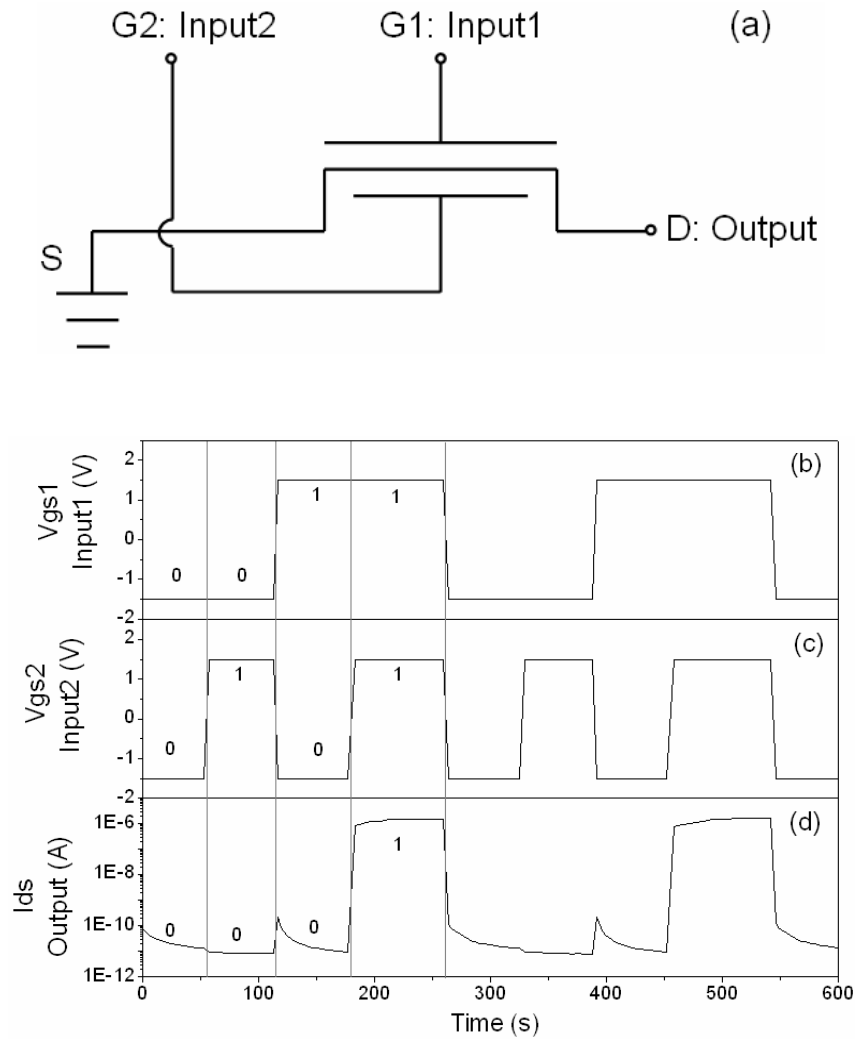

Fig.6 (a) Circuit schematics of a Logic AND applications achieved from our laser patterned dual in-plane-gate (DG) TFT. (b) Gives the input trace applied to G1 as input 1. (c) Gives the input trace applied to G2 as input 2. (HIGH=1.5V as "1", LOW=-1.5V as "0") (d) Shows the $I_{ds}$ output trace. (HIGH>0.7μA as "1", LOW<0.1nA as "0").

Fig.6 shows AND logic operation using a DG TFT, where the logic states ("1", 1.5V; or "0", -1.5V) are input directly and independently to each of the two gates, seen the schematic circuit for inputs 1 and 2. High state ("1", 1.5V) or low state ("0", -1.5V) inputs are applied in sequence through a source measurement units (SMU). The drain current monitored are also detected as output through SMU ("1": $I_{ds}$ above 0.1μA; or "0": $I_{ds}$ below 0.01μA). Only when both of inputs 1 and 2 are HIGH with DG mode, the device drain current is High (>0.7μA). When either or both inputs are LOW, the device drain current is LOW (<0.01nA). The achieved modulation is over four orders of magnitude and the operation of this logic is robust. Though there are some small noises of below 0.1nA, the modulation is still over three orders of magnitude.

## 4. Conclusions

In summary, simple laser scribing process was successfully implemented in fabricating junctionless low-voltage oxide-based TFT arrays with a dual-in-plane gate figure. Robust AND gate logic function was experimentally demonstrated. Such dual in-plane-gate TFTs patterned by laser scribing technology have promising application for lost-cost thin-film electronics.

**This work was supported by the financial supports from the National Natural Science Foundation of China (11174300, 11104288) and the Ningbo Natural Science Foundation Foundation (2011A610202).**

# References

[1] C.H.Park, K.H.Lee, M.S.Oh, K.Lee, S.Im, B.H.Lee, M.M.Sung, IEEE Electron Device Lett., 30(2009)30;

[2] C.W.Lee, A.Afzalian, N.D.Akhavan, R.Yan, I.Ferain, J.P.Colinge, Appl.Phys.Lett., 94(2009)053511;

[3] J.-P.Colinge, C.W.Lee, A.Afzalian, N.D.Akhavan, R.Yan, I.Ferain, P.Razavi, B.ONeill, A.Blake, M.White, A.-M.Kelleher, B.McCarthy, R.Murphy, Nat.Nanotechnol. 5 (2010)225.;

[4] C-J.Su, T-I.Tsai, Y-L.Liou, Z-M.Lin, H-C.Lin, T-S.Chao, IEEE Electron Device Lett., 32(2011)521;

[5] J.Jiang, J.Sun, W.Dou, B.Zhou, Q.Wan, Appl.Phys.Lett., 99(2011)193502;

[6] J.Jiang, J.Sun, W.Dou, Q.Wan, IEEE Electron Device Lett., 33(2012)65;

[7] C.H.Lee, S.J.Kim, Y.S.Oh, M.Y.Kim, Y.J.Yoon, H.S.Lee, J.Appl.Phys. 108(2010)102814;

[8] H.Shin, B.Sim, M.Lee, Optics and lasers in Engineering, 48(2010)816;

[9] H.Lee, H.Shin, Y.Jeong, J.Moon, M.Lee, Appl.Phys.Lett., 95(2009)071104;

[10] 10 Y.Zhou, Q.L.Bao, B.Varghese, L.A.L.Tang, C.K.Tan, C-H.Sow, K.P.Loh, Adv.Mater. 22(2010)67;

[11] J.Jiang, J.Sun, L.Q.Zhu, G.D.Wu, Q.Wan, Appl.Phys.Lett. 99(2011)113504;

[12] A.X.Lu, J.Sun, J.Jiang, Q.Wan, IEEE Electron Device Lett., 31(2010)1137;

[13] H.Yuan, H.Shimotani, A.Tsukazaki, A.Ohtomo, M.Kawasaki, Y.Iwasa,

Adv.Funct.Mater., 19(2009)1046;